# Unveiling Real Triple Degeneracies in Crystals: Exploring Link and Compound Structures

Wenwen Liu[1]†, Hanyu Wang[2]†, Biao Yang[2]*, Shuang Zhang[1]*

[1]Department of Physics, The University of Hong Kong, Hong Kong, China

[2]College of Advanced Interdisciplinary Studies, National University of Defense Technology, Changsha, Hunan 410073, China

*Corresponding authors. Email: yangbiaocam@nudt.edu.cn; shuzhang@hku.hk;

†These authors contributed equally to this work.

**Abstract**

**With their non-Abelian topological charges, real multi-bandgap systems challenge the conventional topological phase classifications. As the minimal sector of multi-bandgap systems, real triple degeneracies (RTPs), which serves as real "Weyl points", lay the foundation for the research on real topological phases. However, experimental demonstration of physical systems with global band configurations consisting of multiple RTPs in crystals have not been reported. Here we present experimental evidence of RTPs in photonic meta-crystals, characterizing them using the Euler number, and establishing their connection with both Abelian and non-Abelian charges. By considering RTPs as the basic elements, we further propose the concept of a topological compound, akin to a chemical compound, where we find that certain phases are not topologically allowed. The topological classification of RTPs in crystals demonstrated in our work plays a similar role as the "no-go" theorem in Weyl systems.**

The study of topological materials has gained widespread attention due to the fruitful phenomena they exhibit. Much effort has been made to identify the characteristics of the system from a topological perspective[1-8]. Recently, there has been growing interest in topological phases that involve multiple bandgaps, which diverge from the traditional focus on single band gaps[5,9-12]. In these phases, the topological invariants depend not only on the individual bandgaps but also on the interactions between them. In the presence of PT (parity and time-reversal) symmetry, the nontrivial topology of multi-bandgap systems can be described using non-Abelian charges[13-17]. These charges are associated with the nodes formed by different bands and resemble the defects in biaxial nematics[13,18,19]. Interestingly, the nodes formed by adjacent bands can only annihilate along specific paths[14,15,20,21]. Euler number is a tool that can be used to characterize annihilation, which is also the counterpart of the Chern number in real-valued systems[22-26]. Similar to the Chern number that characterizes Weyl points, the Euler number can be employed to describe the RTPs [see Supplemental Material, Sec. I].

RTPs are catching growing attention recently as they serve as the basic unit when constructing the non-Abelian topological phases[23,24,27,28]. In two-dimension, they facilitate the movement of nodes and their non-Abelian charges between different energy gaps[15]. Different from the spin-1 Weyl fermion[29,30], these triple points are protected by the combination of inversion and time-reversal symmetries. In classical acoustics and photonics, the zero-frequency points are usually RTPs in the long wavelength limit, which can be understood using the Nambu-Goldstone theorem and characterized by the Euler number[23,31,32]. However, the connection between the Euler charge of the triple point and one-dimensional non-Abelian/Abelian charges still remains elusive. Additionally, the mutual interactions among triple points and their allowed configurations in periodic crystals are largely unexplored. It is well known that Weyl systems[7] are subject to the no-go theorem, which stipulates that the minimal band construction comprising two bands must contain an equal number of positive and negative Weyl points. However, it is unclear yet whether there exists an equivalent no-go theorem that applies to the minimal band construction for RTPs.

Here, we study the topological origin of RTPs from both the non-Abelian and Abelian aspects, and we further demonstrate the existence of various topological phases/compounds consisting of different configurations of RTPs. In the global model, the RTPs can be classified into two types, according to whether the Euler number can be defined or not. The first type, as depicted in Figs. 1a and b, exhibits a fully gapped third band surrounding the RTP. In contrast, the second type, as shown in Figs. 1c and d, does not have fully gapped bands. For clarity in our subsequent discussion, we refer to the fully gapped phase as type-I RTPs that can be characterized by Euler number, while the latter as type-II RTPs where Euler number cannot be defined. Interestingly, under a given symmetry, only certain global configurations (or compound phases) are admissible, which can be regarded as the counterpart of the no-go theory in the real-valued multi-bandgap system. Finally, we experimentally explore the properties of RTPs in a photonic meta-crystal, providing experimental verification for our theoretical framework.

Without loss of generality, we here focus on the $O_h$ point group, while other $PT$ symmetric point groups can be analyzed similarly. Considering a spinless system, we have $PT = K$, which implies $H(\vec{k}) = H(\vec{k})^*$ under an appropriate basis, i.e. $H$ is real at all momentum. For the minimal three-band model, the three eigenstates are real and perpendicular to each other forming an orthogonal frame. Thus, the underlying topology can be described by the rotation of the frame. When assuming two bandgaps, the configuration space of the three-band Hamiltonian is $M_3 = O(3)/O(1)^3$, where $O(3)$ indicates the orthogonal space of all possible collection of the real eigenstates, while $O(1)$ describes the internal degrees of freedom that leave each eigenstate invariant[13]. Applying the fundamental homotopy group we obtain[13] $\pi_1(M_3) = Q_8$, where $Q_8 = \{+1, \pm i, \pm j, \pm k, -1\}$ with $ij = k, jk = i, ki = j$ and $i^2 = j^2 = k^2 = -1$.

By using a three-dimension irreducible representation, we build the effective

Hamiltonian of the triple point and obtain the nodal line structure as shown in Fig. 1a. The Euler charge of the RTP is 2 according to the Wilson loop calculated along the latitude angle $\theta$ on the grey sphere surrounding the triple point (Fig. 1g-h) [Supplemental Material, Sec. II]. After further lowering the crystalline symmetry, e.g. breaking $O_h$, the underlying topology persists under the continuous transformation. As shown in Fig. 1c, if we introduce a strain along the z direction, the nontrivial triple point at $\vec{k} = 0$ would split into two type-II RTPs. However, if we use a sphere to wrap around both type-II RTPs, the Wilson loop still forms a continuous smooth curve corresponding to the Euler number of 2. Further breaking the rotation symmetry along the z direction, a linked structure is formed, as shown in Fig. 1e. The linked structure is very stable under $PT$ symmetry, i.e. it represents the most stable form of the type-II RTP.

The above process also indicates that the type-I RTP is topologically equivalent to two links with the linking number being equal to the Euler number[24]. This can be briefly explained as follows. As shown in Fig. 1i, we consider two green loops that originate from the same base point and encircle around one of the nodal lines at opposite sides of the sphere. It can be computed that the nodal lines carry the same quaternion charge of $i$, which corresponds to the same orientation of the nodal line. Therefore, the lines threading through the sphere experience a trivial Euler charge or non-link structure. In contrast, the two green circles in Fig. 1j exhibit opposite quaternion charges, which corresponds to the opposite orientations of the arrow direction. This means the lines are all going out from the sphere, which indicates the presence of a link structure inside the sphere. If we examine the orientations of the cyan node ring, we can see that it must be linked twice to ensure self-consistency of orientation. This corresponds to the Euler charge of two (Fig S1). Following the process, the type-I RTP characterized by the Euler charge can be transformed into a linked structure.

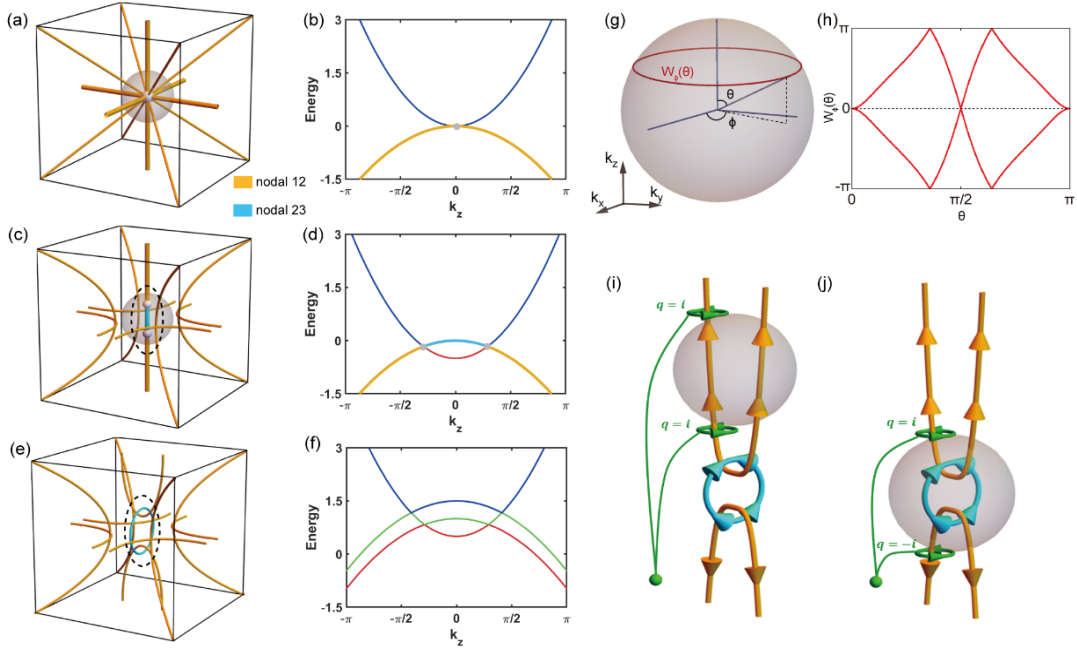

Fig. 1. Topology of RTP (real triple degeneracy). (a) RTP is protected by $O_h$ group. The white point represents the RTP, while the orange lines represent the nodal lines formed by the 1st and 2nd bands. Bold lines indicate quadradic degeneracies from the linear ones which are thinner. (b) The corresponding band structure of (a) along $k_z$ axis. (c) Type-I RTP breaks into two type-II RTPs with lower symmetry. (d) The corresponding band structure of (b) along $k_z$. The cyan line represents the nodal lines formed by the 2nd and 3rd bands. (e) Two type-II RTPs turn out to be a nodal link structure after further lowering the symmetry. (f) The corresponding band structure of (e) along $k_z$ axis. The red, green and blue curves represent the 1st, 2nd and 3rd bands, respectively. (g) Wilson loop path along the circle at certain $\theta$. (h) Wilson loop spectra of the two-band subspace which is symmetric. (i-j) Two different cases are distinguished by non-Abelian charge orientations. The green loops begin from the base point.

For a sphere wrapping around a single type-I RTP, we can analyze the eigenstate distribution to further describe the close connection between the eigenstate-frame rotation and Euler number. An abelian charge is also invoked to facilitate the discussion. The eigenstate of the separate band is always oriented along the radial direction in the vicinity of the nodal point on the sphere, i.e. we could regard the other two eigenstates as the tangent vectors of the sphere. Hence one can focus on the 2D vector singularities on the sphere. Fig. 2a shows the distribution of the eigenstate frame on the sphere wrapping around a type-I RTP (corresponding to Fig. 1a). The inset shows the zoom-in view of the two types of vortices, which indicate two different nodal lines, e.g. quadratic one characterized by non-Abelian charge $-1$, while the linear one characterized by non-Abelian class $\pm i$, respectively. From symmetry, there are 8 singularities with a charge of $\pm i$ and 6 with a charge of $-1$. If we focus on the tangent vector space, the topological classification of these tangent vortices is described by $\pi_1(S^1) = \mathbb{Z}$, which are Abelian topological charges. Here the singularities with non-Abelian charge of $\pm i$ correspond to the Abelian topological charge of $-1$, while those with non-Abelian charge of $-1$ take the Abelian topological charge of $+2$ [33]. In total, there are $-1 \times 8 + 2 \times 6 = +4$ Abelian topological charges, which indicates an Euler number $+2$ for the type-I RTP. It is interesting to see that the indices of the vortices on the sphere follow certain topological rules, i.e. the sum of all the indices is always 2, which is consistent with the Poincare-Hopf theorem[33]. In other words, the singularities of the tangent vector field on the 2-dimensional sphere directly reflect the Euler number.

The eigenstate distribution after breaking rotation symmetry is shown in Fig. 2b (corresponding to Fig. 1c). One can see that the quadratic node splits into two linear nodes, i.e. from non-Abelian charge $-1$ into $\pm i$. However, it should be noted that the new tangent vortices carry an Abelian charge of $+1$. Globally, we still have $-1 \times 8 + 1 \times 8 + 2 \times 2 = +4$, which implies the evolution keeps the underlying topology intact. (Similar analysis can be used in another example which forms zero frequency RTP[32] as shown in Supplemental Material , Sec. IV and Fig. S2).

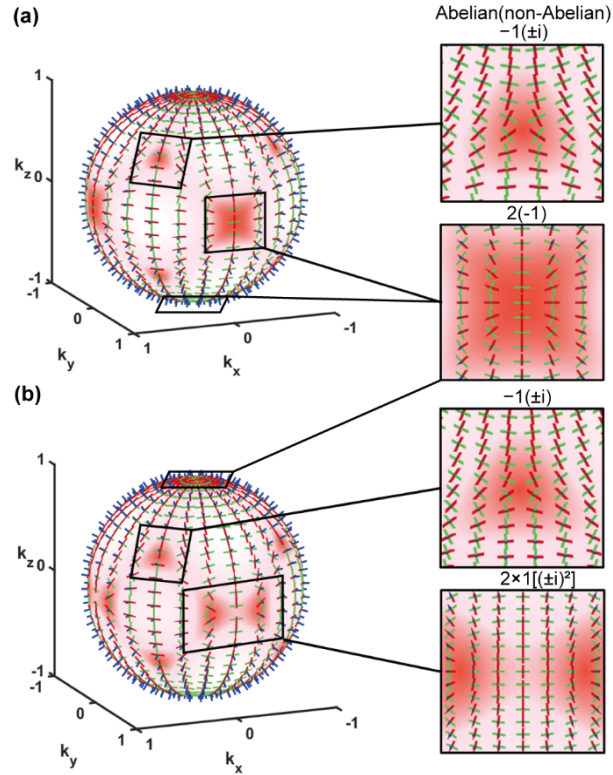

Fig. 2. Abelian (non-Abelian) topological charges of tangent vector vortices. (a) The eigenstate field distribution on a sphere surrounding the type-I RTP. The red, green, and blue line segments represent the real eigenstate of the 1$^{st}$, 2$^{nd}$ and 3$^{rd}$ bands, respectively. The inset shows the zoom-in view of the tangent vector vortices which have different distributions characterized by their corresponding Abelian (non-Abelian) charges. (b) Field distribution after breaking $O_h$ group. One can see that the original tangent vector vortex of non-Abelian charge $-1$ (Abelian topological charge of $+2$) split into two vortices with the non-Abelian class of $\pm i$ (Abelian topological charge of $+1$).

To show the global configurations of the two RTPs, we develop a three-band tight-binding model. We consider $p_x$, $p_y$ and $p_z$ orbitals located on the sites of a 3-dimensional simple cubic lattice which results in $\sigma$ and $\pi$ bonding (see more details in Supplemental Material, Sec. V). In this model, we can have both Type-I and Type-II triple points in one system and they are located at the high symmetry points $\Gamma$ and $R$ (see FBZ) as highlighted by red circles in Fig. 3a. By varying the hopping parameters, we can obtain different configurations of RTPs, e.g., type-I RTP locating on $\Gamma$ and type-II RTP on $R$, they also can change their positions via topological phase transitions. Exhaustive search shows that there are in total three different topological phases, which we call Phase 1, Phase 2 and Phase 3, as marked in Fig. 3b. Phase 3 corresponds to the case that Type-II RTPs occupy both $\Gamma$ and $R$. The band structure shown in Fig. 3a corresponds to Phase 1.

It is interesting to note that the triple points together with the associated nodal lines behave like a chemistry compound. Specifically, the RTPs can be viewed as atoms and the nodal lines as chemical bonds. For instance, the structure can be compared to cesium chloride, which is composed of a simple cubic lattice with a two-atom basis, corresponding to the two different sites ($\Gamma$ and $R$). For each high symmetry point, there are two options for RTPs, which means altogether there are four configurations. Besides the above three configurations, it is found that the fourth configuration wherein both sites are occupied by type-I RTPs (Fig. 3c) is not allowed by the underlying topology. This can be illustrated by Fig. 3c, wherein the nodal line with Abelian charge -1 connecting the two Type-I RTPs ($\Gamma$ and $R$) should contribute the same Euler number to both RTPs. Hence, if one calculates the Euler number on each magenta sphere wrapping around one of the RTPs, the result is equal to 2, with the Abelian charge given by $6 \times 2 - 8 \times 1 = 4$. However, if we use a closed manifold to wrap both $\Gamma$ and $R$, i.e. by smoothly merging the two magenta spheres into a single one in Fig. 3c, the nodal line connecting the two RTPs would not contribute to the Euler number. In this case, the calculation of Euler number is $6 \times 2 - 2 \times (8 - 1) \times 1 = 10$. The result shows that the overall Euler number is inconsistent with adding two individual type-I RTP Euler numbers. In addition, the Euler number equals 5 in this case which violates the fact that it should be even, i.e. $2\mathbb{Z}$. Thus, the no-go theory for RTPs in PT-symmetric systems prohibits the existence of two type-I RTPs linked by a nodal line.

To better understand the classification, we introduce the concept of a quotient graph. The equivalence relation ~ of the system is constrained by space symmetry. Because the point group $O_h$ has $8C_3$ and $6C_4$ symmetries, the eight lines connecting $\Gamma$ and $R$ are equivalent to each other, so are the six lines bridging $\Gamma - \Gamma$ ($R - R$). Therefore, these cases can be simplified into three quotient graphs after eliminating the equivalent classes, as shown in Fig. 3b, wherein different phases are indicated by different colors, and the boundaries between them indicate the topological phase transitions. These transition states at the boundaries consist of triple degenerate lines along the high symmetry lines bridging the points $\Gamma$ and $R$. For example, in the transition from Phase 1 to Phase 2, the triple degenerate lines appear along the diagonals of FBZ, while from Phase 1/2 to Phase 3 the lines appear along the three orthogonal directions, as shown in Fig. 3b. Double degenerate surfaces are also observed in these transition phases, which are plotted in Fig. S4. Note that the triple degenerate lines at the transition states are not robust, in a way similar to the three-dimensional Dirac points lying at the boundary between normal insulators and topological insulators.

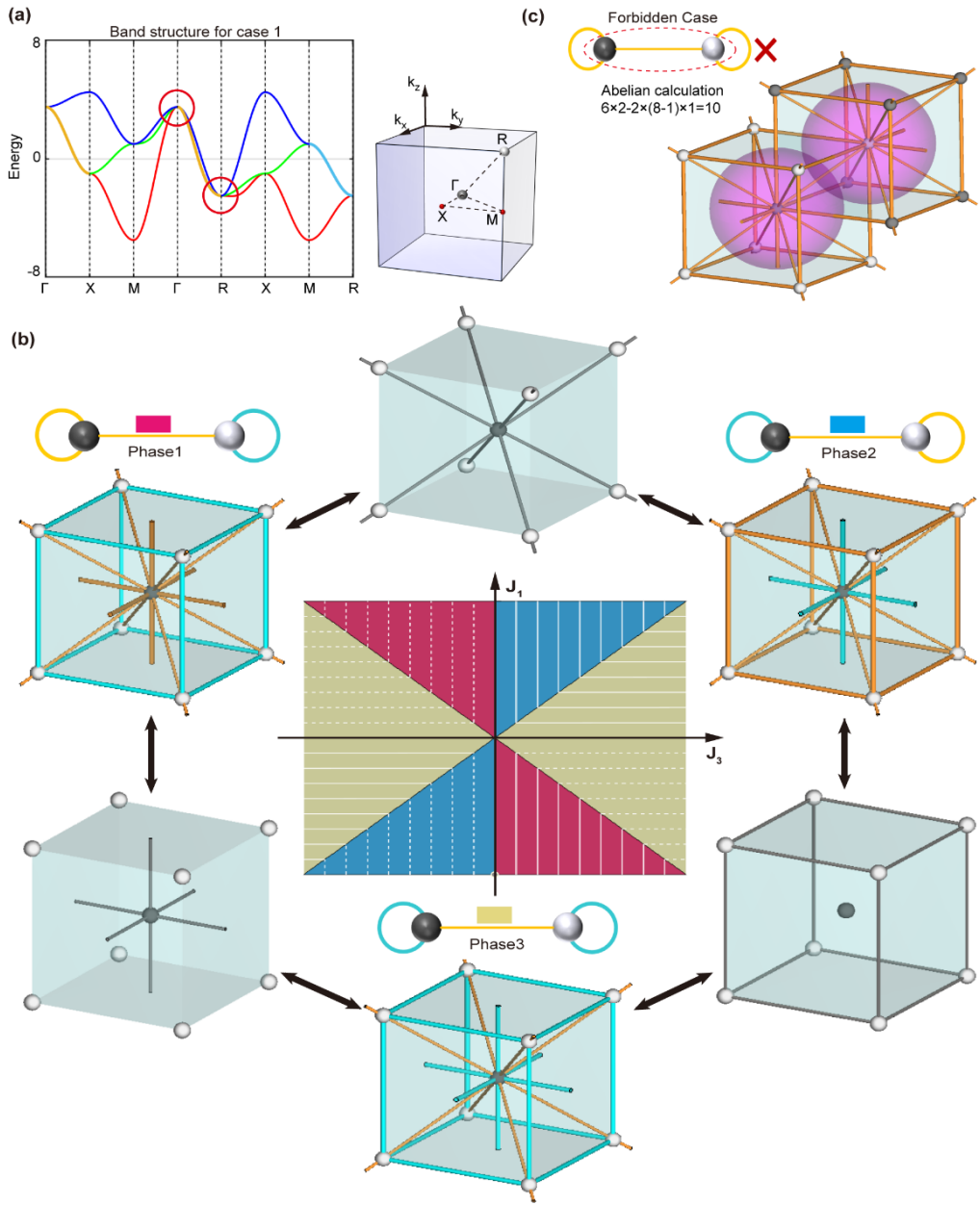

Fig. 3. Topological compound: classification of RTPs in crystals. (a) Band structure of the tight-binding model and first Brillouin zone (FBZ). (b) Topologically admissible cases and their intermediate states. Different colors represent distinct cases and the boundaries are topological phase transitions. The insets show the corresponding quotient graph of the different cases. The orange (cyan) color represents the nodal line formed by the 1st and 2nd (2nd and 3rd) bands. The grey color represents the triple degenerate points and lines. (c) The topologically forbidden phase in the topological compound.

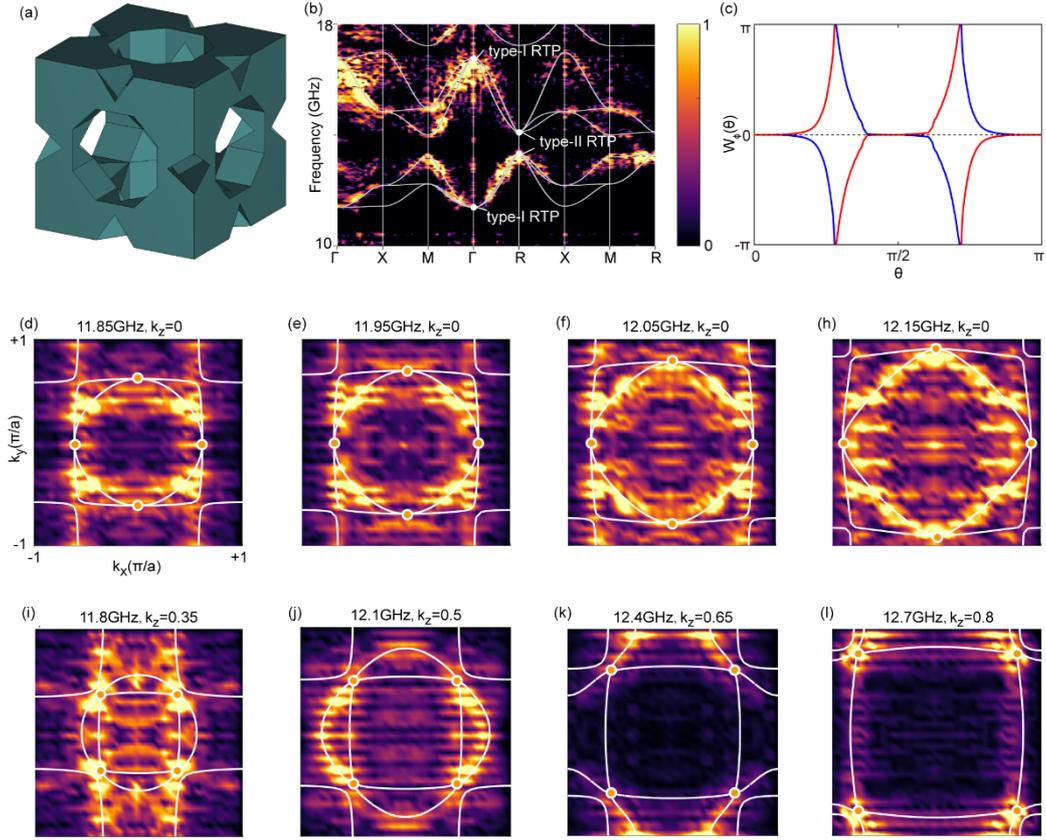

Fig. 4. Experimental observation of real triple points in the photonic crystal. (a) The unit cell of photonic meta-crystal possessing $O_h$ point group. (b) Simulated and mapped band structure along high symmetry lines. (c) Wilson loop calculation of the photonic meta-crystal. (d-l) Equal-frequency contours of the bulk band with varying frequency and $k_z$. The orange points mark the nodes formed by different bands.

For the experimental implementation of RTPs, we design a photonic meta-crystal as shown in Fig. 4a, which belongs to the simple cubic lattice described by $O_h$ point group. The period is $p = 20 \, mm$, other geometry details are given in the Supplemental Material, Fig. S5. We plot the band structure along high symmetry lines in Fig. 4b. With the complete band gap between the 3$^{rd}$ and 4$^{th}$ bands, the six bands are separated into two sets, with each set forming a three-band subspace. The distributions of the RTPs in these two subspaces are similar, as they are both categorized as Phase 1, where RTPs at $\Gamma$ and $R$ correspond to type-I and type-II, respectively. Figure 4c shows the Wilson loop calculation, which indicates a total winding number of 4 and thus the Euler number of

2 for the type-I RTP located at $\Gamma$.

Both the measured and simulated band structures along high symmetry lines are shown in Fig. 4b, which show good agreement with each other. The mapped iso-frequency contours of bulk bands at different frequencies are shown in Fig. 4d-h (the results for three upper bands are shown in Fig. S6), which exhibit nodal degeneracies marked by the orange points. At the $k_z = 0$ plane, it is observed that the nodes shift along $k_x$ and $k_y$ axis with increasing frequency, confirming the existence of nodal lines aligned with the $k_x$ and $k_y$ axis. In contrast, when $k_z \neq 0$, the nodes shift along the diagonal directions with both increasing frequency and $k_z$, as shown in Fig. 4i-l, confirming the existence of nodal lines along the body diagonals of the cubic Brilloun zone.

To conclude, we have experimentally observed the RTPs in photonic meta-crystals and we discovered an analogy between the configuration of RTPs in the momentum space and that of atoms in chemical compounds to describe their properties. By establishing a link between the Euler number and Abelian/non-Abelian charges, we provided a comprehensive understanding of the RTPs. We identified two types of RTPs within a single system and presented various configurations based on different combinations of the RTPs. We also showed that certain compound configurations were forbidden based on a new no-go theory in real-valued systems. While complex systems have been extensively studied, our work introduces a new platform for investigating real-valued topological compounds and their corresponding phase transitions.